\documentclass[twocolumn,pra,showpacs,preprintnumbers,amsmath,amssymb,superscriptaddress]{revtex4}

\usepackage{graphicx}% Include figure files
\usepackage{dcolumn}% Align table columns on decimal point
\usepackage{bm}% bold math
\begin{document}

%\preprint{APS/123-QED}

\title{Three-Component Fermionic Atoms with Repulsive Interaction in Optical Lattices}% Force line breaks with \\

\author{Shin-ya Miyatake}
\affiliation{Department of Applied Physics, Osaka University, Suita, Osaka 565-0871, Japan} 
\author{Kensuke Inaba}%
 \affiliation{NTT Basic Research Laboratories, NTT Corporation, Atsugi 243-0198, Japan and  CREST, JST, Chiyoda-ku, Tokyo 102-0075, Japan}

\author{Sei-ichiro Suga}
\affiliation{Department of Applied Physics, Osaka University, Suita, Osaka 565-0871, Japan}
\affiliation{Department of Materials Science and Chemistry, University of Hyogo, Himeji 671-2280, Japan}

\date{\today}% It is always \today, today,

\begin{abstract}
We investigate three-component (colors) repulsive fermionic atoms in optical lattices using the dynamical mean field theory. 
Depending on the anisotropy of the repulsive interactions, either a color density-wave state or a color selective staggered state appears at half filling. 
In the former state, pairs of atoms with two of the three colors and atoms with the third color occupy different sites alternately. In the latter state, atoms with two of the three colors occupy different sites alternately and atoms with the third color are itinerant throughout the system. When the interactions are isotropic, both states are degenerate. We discuss the results using an effective model. 
\end{abstract}

\maketitle

%%%%%%%%%%%%%%%%%%%%%%%%%%%%%%%%%%%%%%%%%%%%%%%%%%%%%%%%%%%%%%%
%\section{INTRODUCTION}
%%%%%%%%%%%%%%%%%%%%%%%%%%%%%%%%%%%%%%%%%%%%%%%%%%%%%%%%%%%%%%%
Since the first realization of degenerate $^{40}$K fermionic atoms \cite{Jin99}, cold fermionic atoms have been studied extensively. 
By using a Feshbach resonance \cite{Chen2005}, both interaction strength and its sign can be controlled. In addition, optical lattices formed by standing waves of light provide us with diverse interaction configurations. Combining these experimental techniques, fascinating aspects of many-body effects have been revealed. 
A superfluidity of pairs of fermionic atoms was observed for $^{6}$Li fermionic atoms with an attractive interaction \cite{Chin2006}. 
By increasing the depth of the optical lattice potential near the Feshbach resonance, a superfluid-insulator transition was also observed. 
The Mott insulating state was realized for $^{40}$K fermionic atoms with repulsive interaction by appropriately tuning the ratio between the interaction and the kinetic energy \cite{Jordens2008,Schneider2008}. 
Research on cold fermionic atoms has been extended to topics that are not found in ordinary condensed matter physics.

Recently, a balanced population of $^6$Li fermionic atoms with three kinds of internal degrees of freedom was successfully created \cite{ott,Huckans}. The observed anomalous increase in the three-body loss was discussed in connection with the Efimov trimer state \cite{Braaten,Naidon,Floe}. 
For the attractive three-component (colors) fermionic atoms in optical lattices, it was shown theoretically that atoms with two of the three colors form Cooper pairs, yielding a color superfluid (CSF) \cite{Rapp2007,Rapp2008}. As the strength of the attractive interaction increases, there is a quantum phase transition from the CSF state to the trionic state, where three atoms with different colors form singlet bound states \cite{Rapp2007,Rapp2008}. 
Finite-temperature properties including the phase diagram of the Fermi liquid state, the CSF state, and the trionic state have been investigated \cite{Inaba}.

Novel ordered states of repulsive $N$-component cold fermionic atoms have been investigated as regards isotropic interactions on a square lattice \cite{Honer2004}. It was shown that for $N>6$ the staggered flux state appears at half filling. By contrast, for $N<6$, the density-wave (DW) state appears at half filling, where, e.g. for $N=3$, pairs of atoms with two of the three colors and atoms with the third color occupy different sites alternately (color DW state). 
In real systems, the interactions are not necessarily isotropic. However, our knowledge of the effects of the anisotropic interactions is still insufficient. This motivates us to investigate the effects of the  anisotropy of the repulsive interactions.

%%%%%%%%%%%%%%%%%%%%%%%%%%%%%%%%%%%%%%%%%%%%%%%%%%%%%%%%%%%%%%%
%\section{MODEL AND METHOD}
%%%%%%%%%%%%%%%%%%%%%%%%%%%%%%%%%%%%%%%%%%%%%%%%%%%%%%%%%%%%%%%
In this paper, we investigate three-component fermionic atoms with anisotropic repulsive interactions in optical lattices. 
We show that, depending on the anisotropy of the interaction, a color DW state and a color-selective ``antiferromagnetic" (CSAF) state appear at half filling, where atoms with two of the three colors occupy different sites alternately and the third color atoms are itinerant throughout the system. For the isotropic interactions, we find that both states are degenerate.

In accordance with the conventional model for cold atoms in optical lattices \cite{Jaksch}, we set the nearest neighbor hopping and the on-site interaction between atoms with different colors. The low-energy properties can be described by the following Hamiltonian, 
%****************************************************************
\begin{eqnarray}
{\cal H}&=-&\sum_{<i,j>}\sum_{\alpha=1}^{3}
       \left( t+\mu_\alpha \delta_{i,j} \right)a^\dag_{i\alpha} a_{j\alpha} 
   \nonumber \\
   &+& \frac{1}{2}\sum_{i}\sum_{\alpha\not=\beta} 
       U_{\alpha\beta} n_{i \alpha} n_{i \beta}
\label{eq_model},
\end{eqnarray}
%****************************************************************
where the subscript $<i,j>$ is the summation over the nearest neighbor sites, $a^\dag_{i\alpha} (a_{i\alpha})$ is the fermionic creation (annihilation) operator for the state with color $\alpha$ in the $i$th site and $n_{i\alpha}=a^\dag_{i\alpha}a_{i\alpha}$. 
$t$ denotes the hopping integral, 
$\mu_\alpha$ is the chemical potential for the atom with color $\alpha$, and 
$U_{\alpha\beta}$ is the repulsive interaction between two atoms with colors $\alpha$ and $\beta$. 
We assume a homogeneous optical lattice and neglect the confinement potential for a first approximation.

To investigate quantum phase transitions and the ground state properties, we employ the dynamical mean field theory (DMFT) \cite{Georges1996}. 
In DMFT, the lattice model is mapped onto an effective impurity model connected dynamically to a heat bath. The Green's function is obtained via the self-consistent solution of this impurity problem. 
This retains nontrivial quantum fluctuations missing in conventional mean-field theories. 
We apply here the two-site DMFT method to obtain this self-consistent solution  \cite{Potthoff2001}, which allows us to study the DW states and the Mott insulating states of lattice fermions \cite{Higashiyama2009}.

%*******************************************
\begin{figure}[tb]
\begin{center}
\includegraphics[scale=0.35]{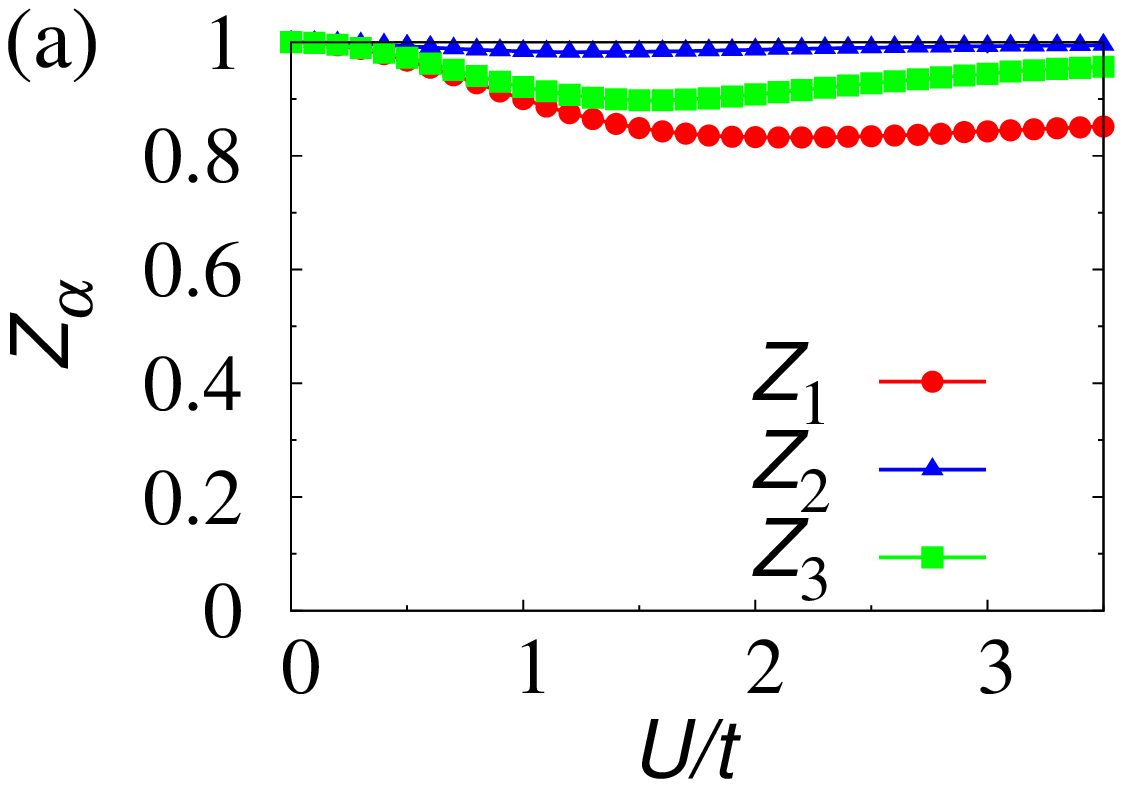}
\includegraphics[scale=0.35]{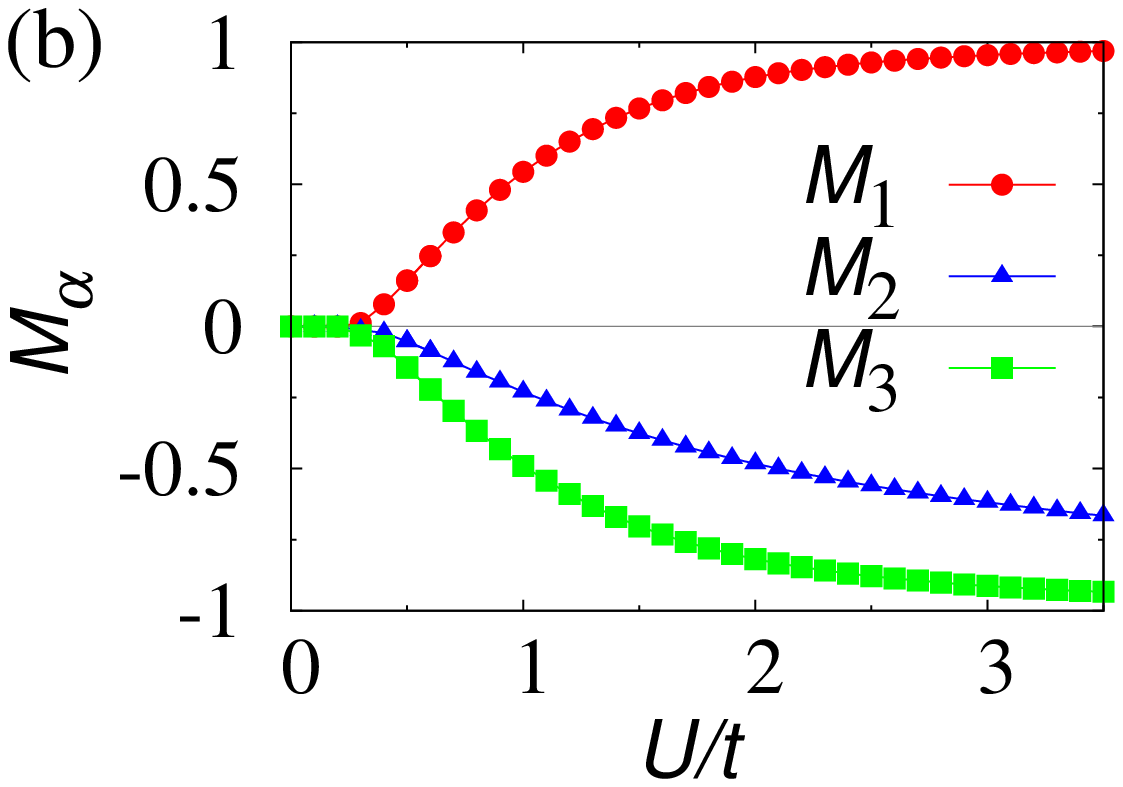}
\caption{(Color Online) (a) Quasiparticle weight $Z_{\alpha}$ and (b) difference between atom numbers of sublattices with color $\alpha$ per site, $M_{\alpha}$, for $U''/U=2$ and $U'/U=0.6$ as functions of $U/t$. 
}
\label{fig1}
\end{center}
\end{figure}
%*******************************************

To examine the color DW state, we divide the bipartite lattice into two sublattices \cite{Georges1996,Chitra}. 
In this procedure, the local Green's function has the following form: 
%****************************************************************
%\begin{widetext}
\begin{eqnarray}
{\hat{G}_{\alpha}^{}(\omega)} &=&
\int d \varepsilon
      D(\varepsilon) \left[{\hat{g}_{\alpha}^{-1}
        (\varepsilon,\omega)-\hat{\Sigma}_\alpha(\omega)}\right]^{-1}, \\
{\hat{g}_{\alpha}^{-1}(\varepsilon,\omega)} &=&
\left(
 \begin{array}{ccc}
  \omega+\mu_{\alpha} & - \varepsilon \\
  - \varepsilon & \omega+\mu_{\alpha}
 \end{array}
\right), \\
\hat{\Sigma}_\alpha(\omega)&=&
\left(
 \begin{array}{ccc}
\Sigma_{A,\alpha}(\omega) & 0 \\
0&\Sigma_{B,\alpha}(\omega)
 \end{array}
\right),
\label{green}
\end{eqnarray}
%\end{widetext} 
%****************************************************************
where $D(\varepsilon)$ is the density of states (DOS). 
We use a semicircular DOS obtained for the infinite dimensional bipartite Bethe lattice, $D( \varepsilon)=4/(\pi W_{})\sqrt{1-4( \varepsilon/W_{})^{2}}$, where $W(=4t)$ is the bandwidth. Accordingly, qualitative properties of other bipartite lattices such as square and cubic lattices can be discussed using the present results. 
$\Sigma_{A(B),\alpha}(\omega)$ is the self-energy of the atoms with color $\alpha$ for the $A (B)$ sublattice, which can be obtained by solving two effective impurity models. 
The chemical potential is set at $\mu_{\alpha}=(U_{\alpha \beta}+U_{\gamma \alpha})/2$ so that the particle-hole symmetry can be realized for each color, and thus the atoms of each color achieve half filling. 
We set $U_{12}=U$, $U_{23}=U'$, and $U_{31}=U''$.

%%%%%%%%%%%%%%%%%%%%%%%%%%%%%%%%%%%%%%%%%%%%%%%%%%%%%%%%%%%%%%%
%\section{NUMERICAL RESULTS}
%%%%%%%%%%%%%%%%%%%%%%%%%%%%%%%%%%%%%%%%%%%%%%%%%%%%%%%%%%%%%%%
%\subsection{Attractive interaction}
%%%%%%%%%%%%%%%%%%%%%%%%%%%%%%%%%%%%%%%%%%%%%%%%%%%%%%%%%%%%%%%
We calculate the quasiparticle weight $Z_{\alpha}$, which is inversely proportional to the effective mass of the atom with color $\alpha$, and the difference in the atom numbers ($n_{A(B),\alpha}$), $M_{\alpha}=n_{A,\alpha}-n_{B,\alpha}$, between the sublattices $A$ and $B$ with color $\alpha$ per site. 
We also calculate the single-particle excitation spectra (SPES), $\rho_{A(B) \alpha}(\omega)$, defined from the imaginary part of the diagonal components of the $2 \times 2$ Green's function matrix, $-(1/\pi){\rm Im}\hat{G}_{\alpha}(\omega+i\delta)$, with $\delta$ being a small positive number. 
Because of particle-hole symmetry, the relation $\rho_{A \alpha}(\omega)=\rho_{B \alpha}(-\omega)$ is satisfied. 
In the following, the hopping integral $t$ is used in units of energy.

%*******************************************
\begin{figure}[tb]
\begin{center}
\includegraphics[scale=0.4]{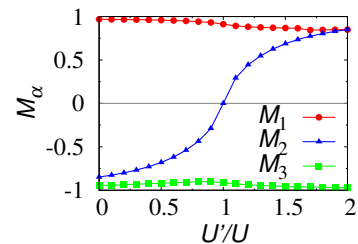}
\caption{(Color Online) Difference between atom numbers of sublattices with color $\alpha$ per site, $M_{\alpha}$, for $U''/U=2$ and $U/t=3$ as functions of $U'/U$. 
}
\label{fig2}
\end{center}
\end{figure}
%*******************************************
%*******************************************
\begin{figure}[tb]
\begin{center}
\includegraphics[scale=0.42]{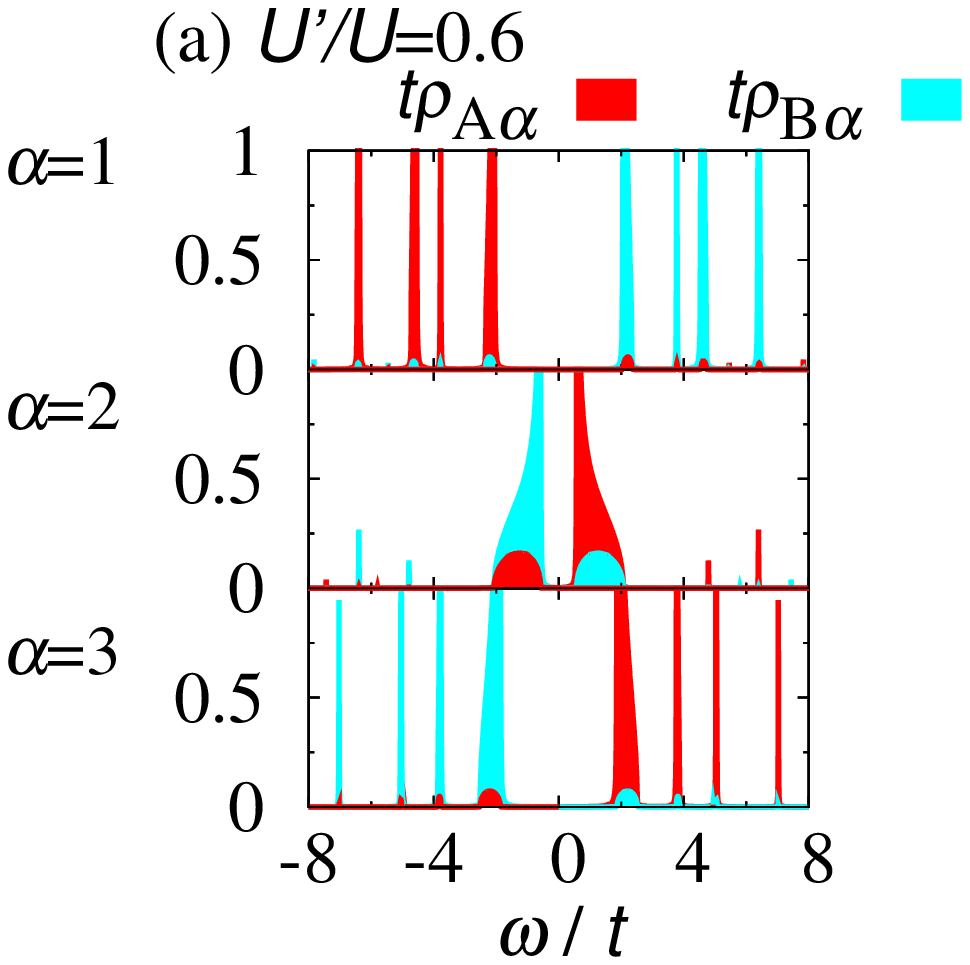}
\includegraphics[scale=0.42]{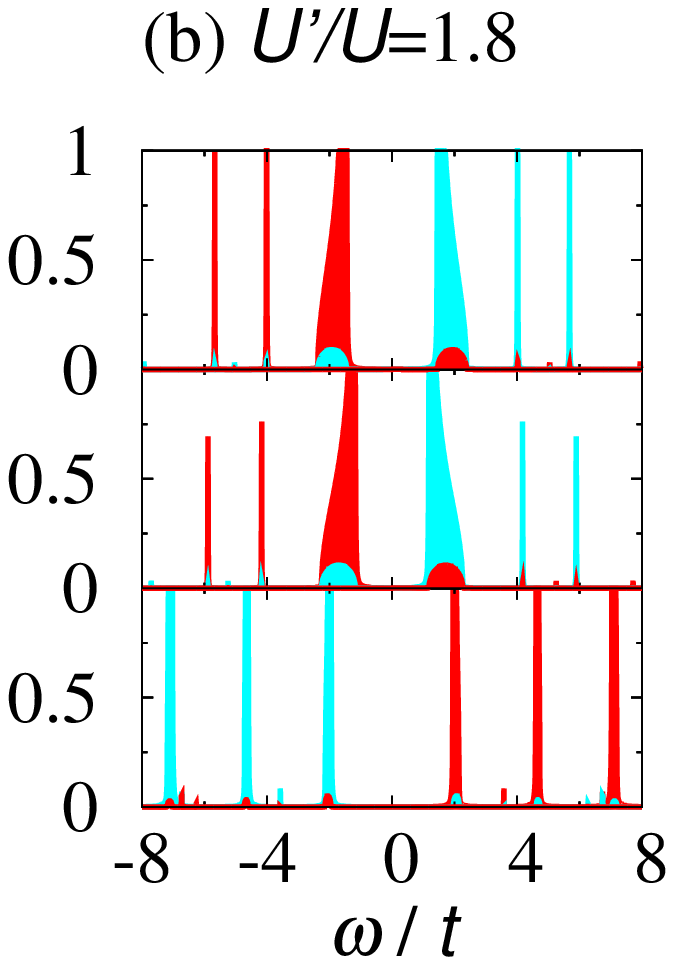}
\caption{(Color Online) Single-particle excitation spectra for (a) $U'/U=0.6$ and (b) $1.8$ by fixing $U''/U=2$ and $U/t=3$. 
%SPES expressed by red (black) and blue (gray) are from $\rho_{A,\alpha}$ and $\rho_{B,\alpha}$, respectively. 
%The hopping integral $t$ is used in units of energy. 
}
\label{fig3}
\end{center}
\end{figure}
%*******************************************
In Fig. \ref{fig1}, we show $Z_{\alpha}$ and $M_{\alpha}$ for $U''/U=2$ and $U'/U=0.6$ as functions of $U/t$. 
As $U/t$ increases, $Z_{1}, Z_{2}$, and $Z_{3}$ first decrease and then maintain nonzero values even in a large $U/t$. 
With increasing $U/t$, $M_{1}$ increases towards unity, and $M_{2}$ and  $M_{3}$ decrease towards minus unity. 
The results indicate that no Mott transition occurs, and sites doubly occupied by atoms with colors 2 and 3 and sites solely occupied by atoms with color 1 appear alternately, and form the color DW state. 
We find that $Z_2 \sim 1$ irrespective of $U/t$, implying a weak renormalization effect, and that $M_{2}$ approaches minus unity very slowly compared with $M_{3}$. The results suggest that the atoms with color 2 occupy the site more flexibly than the atoms with other two colors. 

In Fig. \ref{fig2}, we show $M_{\alpha}$ as functions of $U'/U$ for $U''/U=2$ and $U/t=3$. We find that $M_{1} \sim 1$ and $M_{3} \sim -1$ irrespective of $U'/U$, indicating that the atoms with colors 1 and 3 occupy the alternating sites rigidly. By contrast, $M_{2}$ changes from negative to positive with increasing $U'/U$. 
The results indicate that for $U'/U<1$ the atoms with colors 2 and 3 occupy the same sites to prevent the strong repulsions $U''$ and $U$, while for $U'/U>1$ the atoms with colors 1 and 2 occupy the same sites to prevent the strong repulsions $U''$ and $U'$. 
Close to $U'/U=1$, $M_{2} \sim 0$. In this case, the binding of the color 2 atoms to the color DW state becomes very weak.  
At $U'/U=1$, $M_{2}$ vanishes, suggesting that the atoms with color 2 are itinerant throughout the system. In this case, the CSAF state appears.

We investigate the color DW state in terms of the SPES. 
In Fig. \ref{fig3}, we show the SPES for $U'/U=0.6$ and $1.8$ by fixing $U''/U=2$ and $U/t=3$. We find that the SPES have gaps $(\Delta_{\alpha})$ at the Fermi energy $\omega=0$. 
For $U'/U=0.6$, the gap of color 2 $(\Delta_{2}/t=1.21)$ is much smaller than those of colors 1 and 3 $(\Delta_{1}/t=4.02$ and $\Delta_{3}/t=3.62)$. This is because the atoms with colors 1 and 3 occupy different sites more rigidly to prevent the largest repulsion $U''$ than the atoms with color 2. 
The results are consistent with those for $Z_{\alpha}$ and $M_{\alpha}$ shown in Fig. \ref{fig1}, where $M_{1}=0.95$, $M_{2}=-0.62$, and $M_{3}=-0.91$ for $U''/U=2$, $U'/U=0.6$, and $U/t=3$. 
For $U'/U=1.8$, we obtain the spectral gaps of $\Delta_{1}/t=2.80$, $\Delta_{2}/t=2.25$, and $\Delta_{3}/t=3.82$. No gap is noticeably smaller than the others. The situation is consistent with the color DW order parameters: $M_{1}=0.85$, $M_{2}=0.82$, and $M_{3}=-0.97$ for $U''/U=2$, $U'/U=1.8$, and $U/t=3$.

%*******************************************
\begin{figure}[tb]
\begin{center}
\includegraphics[scale=0.35]{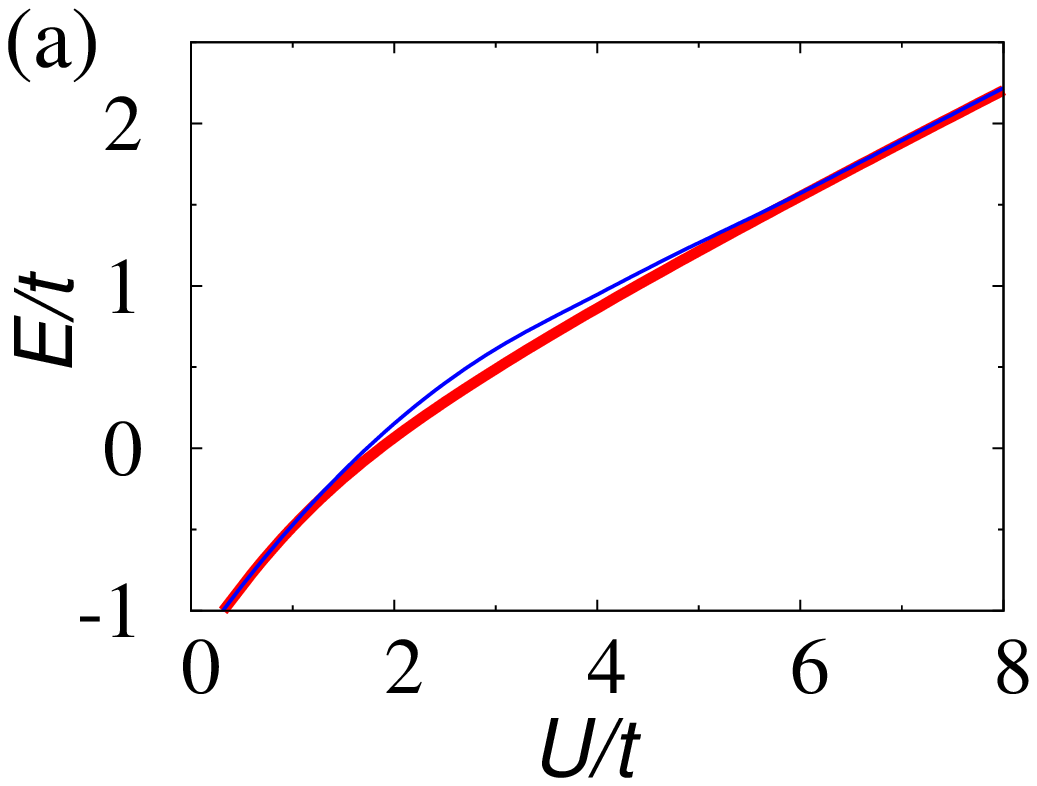}
\includegraphics[scale=0.35]{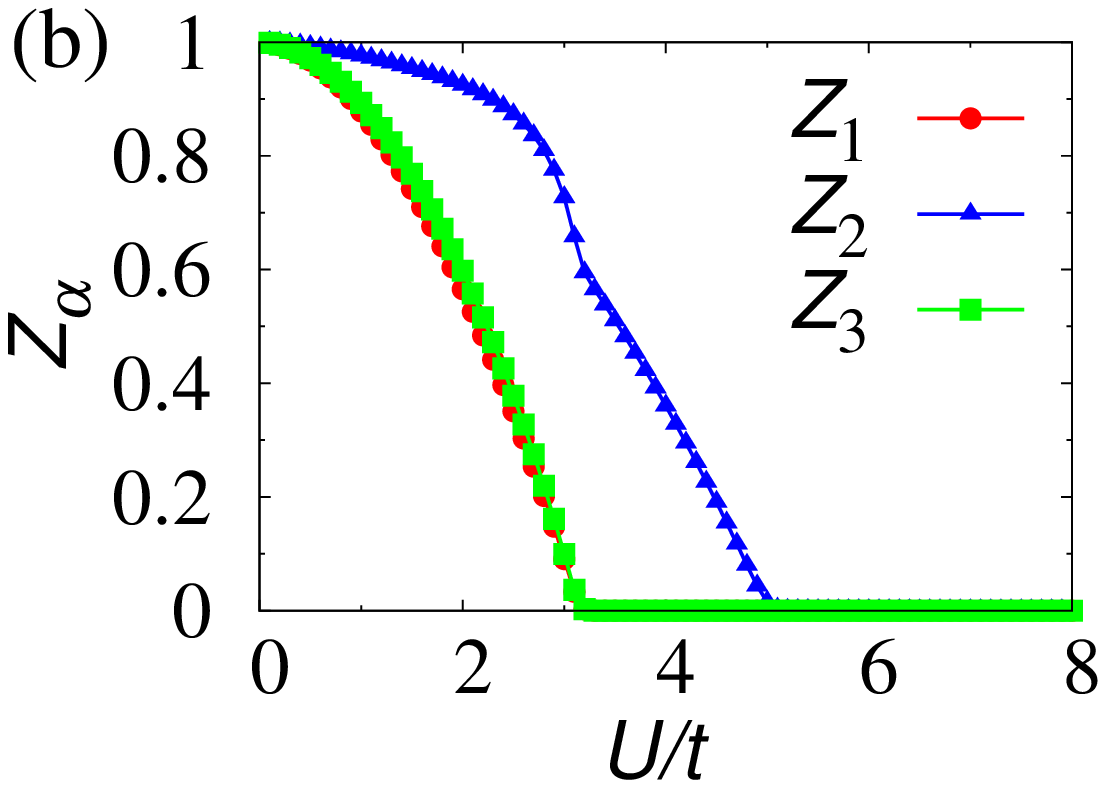}
\caption{(Color Online) (a) Energies and (b) quasiparticle weights for $U''/U=2$ and $U'/U=0.6$ as functions of $U/t$. (a) The thick (red) line represents the energy of the color DW state, while the thin (blue) line represents the energy obtained by the two-site DMFT without dividing the bipartite lattice into the sublattices.  
}
\label{fig4}
\end{center}
\end{figure}
%*******************************************
The results in Figs. \ref{fig1}(a) and \ref{fig3}(a) show that a small spectral gap appears, when one of the three repulsive interactions is particularly strong. 
In this case, fluctuation effects caused, for example, by temperature are considered to have a noticeable influence on the low-energy properties. Owing to thermal fluctuation effects, the atoms corresponding to the small spectral gap may become itinerant. 
To confirm this consideration, we calculate $Z_{\alpha}$ using two-site DMFT without dividing the bipartite lattice into sublattices. In this calculation, the atoms with different colors do not form any staggered state such as the color DW state or the CSAF state. Instead, atoms with a strong repulsive interaction occupy the sites randomly, obeying a Mott transition. 
This calculation allows us to study quasi-stable states that have higher energies than those of the ordered state. In fact, the energy obtained by this calculation is slightly higher than that of the color DW state as shown in Fig. \ref{fig4}(a). 
The quasiparticle weights for the same parameters ($U''/U=2$ and $U'/U=0.6$) as those used for Fig. \ref{fig1} are shown in Fig. \ref{fig4}(b) as functions of $U/t$. 
With increasing $U/t$, $Z_{\alpha}$ decreases monotonically. 
$Z_{1}$ and $Z_{3}$ decrease and become zero at the same critical $U(=U_{c1})$, which is smaller than the critical $U(=U_{c2})$ for $Z_{2}$. Accordingly, for $U_{c1}<U<U_{c2}$, the atoms with colors 1 and 3 exhibit a Mott transition because of the strong interaction $U''$, while the atoms with color 2 behave as a Fermi liquid and are itinerant throughout the system. Using the analogy of the orbital selective Mott transition \cite{OSMT}, this state can be called a color selective Mott transition (CSMT) state. 
For $U_{c2}<U$, the Mott insulator appears. 
It was demonstrated that the Mott insulator appears at commensurate $1/3$  and $2/3$ fillings (one atom and two atoms in each site, respectively) in the isotropic interaction system \cite{Gorelik,Miyatake}. 
We have, in addition, shown that if the interactions are anisotropic the Mott insulator appears even at incommensurate half filling (non-integer 1.5 atoms in each site).  
The present calculation suggests that atoms with a small energy gap in the ground state may become itinerant at finite temperatures. 
In this case, there may be competition between the color DW state, the CSAF state, and the CSMT state depending on the anisotropy of the repulsive interactions. This issue constitutes our future study.

%*******************************************
\begin{figure}[tb]
\begin{center}
\includegraphics[scale=0.35]{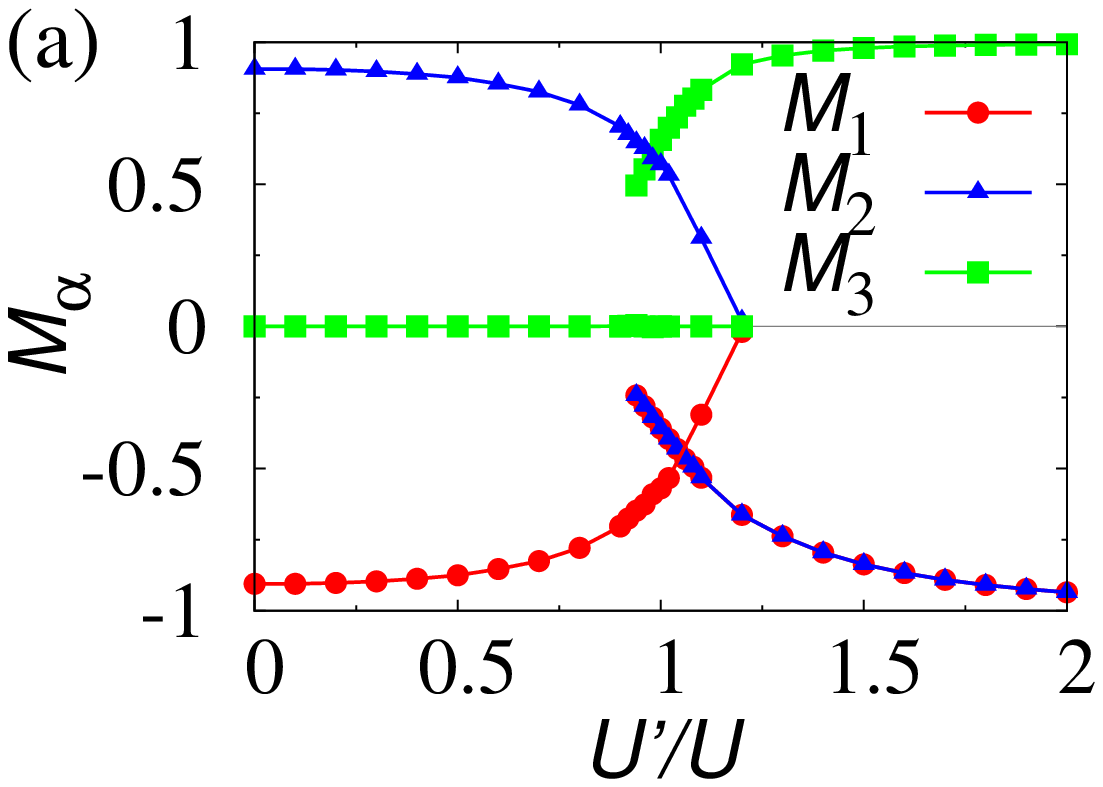}
\includegraphics[scale=0.35]{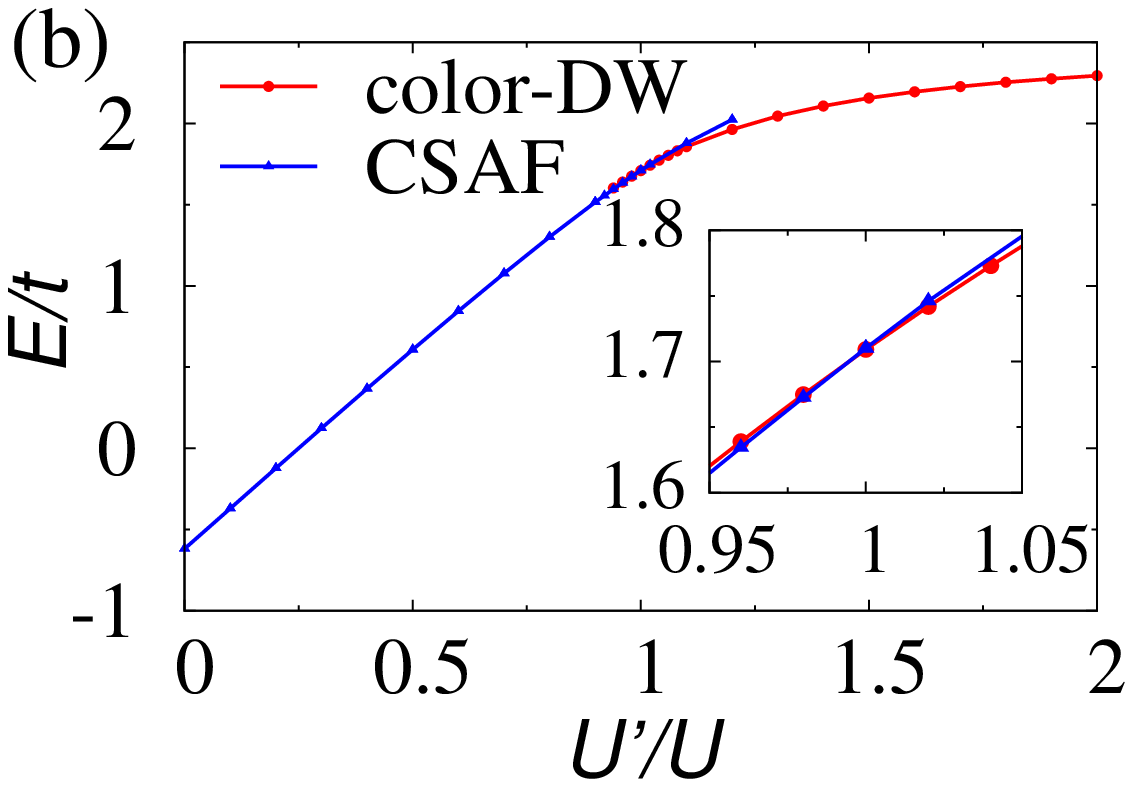}
\caption{(Color Online) (a) Order parameters and (b) energies for $U''=U'$ and $U/t=5$ as functions of $U'/U$. 
}
\label{fig5}
\end{center}
\end{figure}
%*******************************************
We have revealed that the anisotropy of the interactions induces novel quantum phases. We now move our attention close to the isotropic interaction point ($U''=U'=U$), i.e. SU(3) point, which was investigated in Ref. \cite{Honer2004}. 
To this end, we calculate $M_{\alpha}$, and the energies of the color DW state and CSAF state for $U''=U'$ and $U/t=5$, when $U'/U$ varies across the SU(3) point ($U'/U=1)$. 
The results for $M_{\alpha}$ are shown in Fig. \ref{fig5}(a) as functions of $U'/U$. 
For $U'/U<0.94$ the CSAF state appears, since $M_1<0$, $M_2>0$, and $M_3=0$. For $U'/U>1.20$ the color DW state appears, since $M_1<0$, $M_2<0$, and $M_3>0$. 
We find a hysteresis in $0.94<U'/U<1.20$, implying a discontinuous quantum phase transition. 
The results for the energies are shown in Fig. \ref{fig5}(b). 
Although the difference between the energies is very small, the CSAF state is the lowest for $U'/U<1$ and the color DW state is the lowest for $U'/U>1$. At $U'/U=1$, the energies of both states cross, which is a manifestation of a discontinuous quantum phase transition in agreement with the results for $M_{\alpha}$. 
We have shown that the color DW state and the CSAF state are degenerate at the SU(3) point in the infinite-dimensional Bethe lattice. 
As mentioned above, our results can be applied to discuss qualitative properties of a square lattice system. 
In Ref. \cite{Honer2004}, it was claimed that the color DW state is the ground state of the isotropic interaction system on a square lattice at half filling. It may be an interesting issue to investigate this degeneracy in other bipartite lattices, e.g. a cubic lattice.

We discuss the three-component lattice fermions at half filling using an effective model. This effective model is appropriate, when at least one of the three repulsive interactions is very strong. When $U''$ is the strongest, the atoms with colors 1 and 3 occupy completely different sites at half filling, forming localized fermions. Note that the localized fermions do not necessarily form staggered states. 
To derive the effective Hamiltonian, we set $N_f=\frac{1}{2}(n_1-n_3+1)$ and $N_c=n_2$, where $n_{\alpha}=\frac{1}{2}(n_{A,\alpha}+n_{B,\alpha})$. In this procedure, the atoms with color 2 are regarded the spinless fermions. 
To evaluate the interaction between the localized fermions and the spinless fermions, and the chemical potentials of both type of fermions, the following four states are considered: $(N_f, N_c)=(1,1), (1,0), (0,1), (0,0)$. 
The effective Hamiltonian can be obtained as 
%****************************************************************
\begin{eqnarray}
{\cal H}_{eff} = &-& t\sum_{<i,j>}c^\dag_{i}c_{j} 
       - \mu \sum_{i}\left(c^\dag_{i}c_{i}+f^\dag_{i}f_{i} \right)
   \nonumber \\
   &+& (U-U')\sum_{i} c^\dag_{i}c_{i}f^\dag_{i}f_{i}, 
\label{eq_eff}
\end{eqnarray}
%****************************************************************
where $c_i$ and $f_i$ are the annihilation operators of the spinless fermions and the localized fermions, respectively, and  $\mu=\frac{1}{2}(U-U')$ $(>0)$ \cite{comm}. This Hamiltonian is known as a Falicov-Kimball model for symmetric half filling, the ground state of which is known to be an antiferromagnetic DW state \cite{Brandt1989,Georges1996}. In terms of the original model, this ground state is the same as the color DW state. 
For a small $U-U'$, it can be considered that the system consists of spinless fermions loaded in the staggered periodic potential $0$ and $U-U'$ induced by the DW state of the localized fermions, which corresponds to the AF insulating state formed by the color-1 and -3 atoms in the original model. 
The Fermi surface of the spinless fermions is perfectly nested, resulting in a band-insulating state with a small gap corresponding to $\Delta_2(\propto |U-U'|)$ as shown in Fig. \ref{fig3}(a). 
The CSAF state appearing at $U=U'$ can be described by the effective model, if the localized fermions form the staggered DW state even at $U'=U$.

%%%%%%%%%%%%%%%%%%%%%%%%%%%%%%%%%%%%%%%%%%%%%%%%%%%%%%%%%%%%%%%
%\section{SUMMARY}
%%%%%%%%%%%%%%%%%%%%%%%%%%%%%%%%%%%%%%%%%%%%%%%%%%%%%%%%%%%%%%%
In summary, we have investigated three-component repulsive fermionic atoms in optical lattices. Depending on the anisotropy of the repulsive interaction, the color DW state and the CSAF state appear at half filling. Both states are degenerate at the SU(3) isotropic interaction point. 
In recent experiments, it was shown that fermionic ytterbium atoms ($^{173}$Yb) were successfully cooled to form degenerate Fermi gases \cite{Takahashi}. The interaction between $^{173}$Yb atoms is repulsive and its nuclear spin is $I=5/2$ \cite{Kitagawa}. Accordingly, $^{173}$Yb fermionic atoms may be a candidate for realizing and studying three-component repulsive fermionic atoms in optical lattices. We hope that our results contribute to work on the novel ordered states of cold fermionic atoms in optical lattices.

%%%%%%%%%%%%%%%%%%%%%%%%%%%%%%%%%%%%%%%%%%%%%%%%%%%%%%%%%%%%%%%
%\section{ACKNOWLEDGEMENTS}
%%%%%%%%%%%%%%%%%%%%%%%%%%%%%%%%%%%%%%%%%%%%%%%%%%%%%%%%%%%%%%%
We thank Y. Takahashi and M. Yamashita for useful comments and valuable discussions. 
Some of the numerical computations were undertaken at the Supercomputer Center at ISSP, University of Tokyo. 
This work was supported by Grant-in-Aids for Scientific Research (C)
(No. 20540390) from the Japan Society for the Promotion of Science and on Innovative Areas (No. 21104514) from the Ministry of Education, Culture, Sports, Science and Technology.

%\newpage %Just because of unusual number of tables stacked at end

\begin{thebibliography}{00}
\bibitem{comm} If we set $N_f=\frac{1}{2}(-n_1+n_3+1)$, the coupling constant of the effective Hamiltonian is $U'-U$ and $\mu=\frac{1}{2}(U'-U)$. We should select a definition of $N_f$ that allows us to obtain the positive coupling constant and chemical potential.
\end{thebibliography}
%\reviseComment{Refの[6]-[11]と[15]がAPSのフォーマットになっていません。}
%%%%%%%%%%%%%%%%%%%%%%%%%%%%%%%%%%%%%%%%%%%%%%%%%%%%%%

\end{document}